# The role of hydrostatic stress in determining the bandgap of InN epilayers

Abdul Kadir[1,*], Tapas Ganguli[2], Ravi Kumar[2], M.R. Gokhale[1], A.P. Shah[1], Sandip Ghosh[1], B.M. Arora[1] and Arnab Bhattacharya[1]

[1]Tata Institute of Fundamental Research, Homi Bhabha Road, Mumbai 400005, India
[2]Raja Ramanna Center for Advanced Technology, Indore 425013, India

*Email: abdulkadir@tifr.res.in

May 17, 2007

**Abstract**
We establish a correlation between the internal stress in InN epilayers and their optical properties such as the measured absorption band edge and photoluminescence emission wavelength. By a careful evaluation of the lattice constants of InN epilayers grown on *c*-plane sapphire substrates under various conditions by metalorganic vapor phase epitaxy we find that the films are under primarily hydrostatic stress. This results in a shift in the band edge to higher energy. The effect is significant, and may be responsible for some of the variations in InN bandgap reported in the literature.

Among the group III-Nitride semiconductors, InN has been extensively investigated in recent years due to its excellent electrical properties predicted theoretically and for its potential device applications. Research on this material has been further intensified by the controversy in the bandgap value. There has been a significant improvement in the quality of InN epilayers, and a large body of recent work suggests that the bandgap of InN is of the order of 0.7eV-0.8eV, e.g. reviews[1,2], which is much smaller than the previously reported value of 1.8-1.9eV (Ref. 3). There is, however, still uncertainty on the cause of variations in InN bandgap reported by different groups and these differences in bandgap have been attributed to factors such as Moss-Burstein shift[4,5], oxygen alloying[5], presence of metallic In and Mie scattering from such In droplets[6], presence of trapping levels[7] and quantum size effects[8]. Compared to the large volume of literature on the bandgap controversy in InN, there is relatively little published data that discusses any systematic dependence of the measured bandgap of InN to the lattice parameter(s) of InN. The scatter in the reported values of the "*c*" and "*a*" lattice constants of InN is quite large and has almost always been attributed to the inbuilt strain (biaxial) in the InN layers due to the effects of the substrate/buffer[9]. There are only a few reports on the presence of hydrostatic strain in InN films[10]. We have earlier reported a detailed study of the growth parameter space for InN in a close-coupled showerhead metalorganic vapor phase epitaxy (MOVPE) system, examining the effects of V/III ratio, temperature, reactor pressure, precursor flux, nitridation/annealing temperature etc.[11] As a result of this study covering over 40 growth experiments we have access to a range of InN samples, a systematic analysis of whose lattice parameters allows us to evaluate the stress in the epilayer and compare with the corresponding absorption/photoluminescence measurements after correcting for effects due to the carrier concentration. In this letter we address the role of hydrostatic stress in determining the bandgap of such MOVPE-grown InN epilayers.

All the InN epilayers studied in this work were deposited by MOVPE on 2" *c*-plane sapphire substrates in a 3x2" close-coupled-showerhead reactor (Thomas Swan) using trimethylindium (TMIn) and ammonia ($NH_3$) as precursors with nitrogen as the carrier gas. The details of the growth have been discussed previously[11]. The InN films were structurally characterized by high-resolution X-ray diffraction (HRXRD) on a PANalytical X-pert MRD system with a Hybrid 4-bounce monochromator at the input having a divergence of ~ 20 arc seconds. This system has also been used for the determination of the lattice parameters for the InN films based on the method described in Ref. 12. The corrections due to the centering of the sample on the goniometer have been made using the extrapolation formula: $\Delta d/d = -(D/r)\cos^2\theta/\sin\theta$, where d is the diffraction plane spacing, $\Delta d$ is the difference between the measured and the actual plane spacing, D is the displacement of the sample with respect to the goniometer axis in the equatorial plane, and r is the radius of the goniometer. Other corrections like those due to refractive index, Lorentz-polarization and absorption together lead to an estimated lattice parameter inaccuracy of about $1 \times 10^{-5}$Å. As the range of the lattice parameters observed in this work are greater than 1%, such corrections can be neglected. Room/low temperature photoluminescence (PL) measured using a 0.67m monochromator and $Ar^+$ ion laser excitation, and absorption measurements (on back-side polished samples) using a Cary 5000 UV/VIS/NIR spectrophotometer were used to estimate the bandgap. Hall measurements in the Van-der-Pauw geometry were used to determine the carrier concentration and the carrier mobility.

The strain in the InN layers is evaluated from the observed value of the strained and the unstrained lattice parameters (5.7064Å and 3.5376Å for "*c*" and "*a*" respectively[13]). The stress tensor components in the samples are then evaluated from the values of the compliance tensor for InN using the data in Ref. 14.

Fig. 1a shows ω-2θ x-ray scans for 4 representative 0.2μm-thick InN layers deposited at 300 Torr and 530°C on sapphire with a GaN buffer, at varying V/III ratios. The inset shows details of the InN (0002) peak which varies for different layers. Broadly, the 2θ positions fall

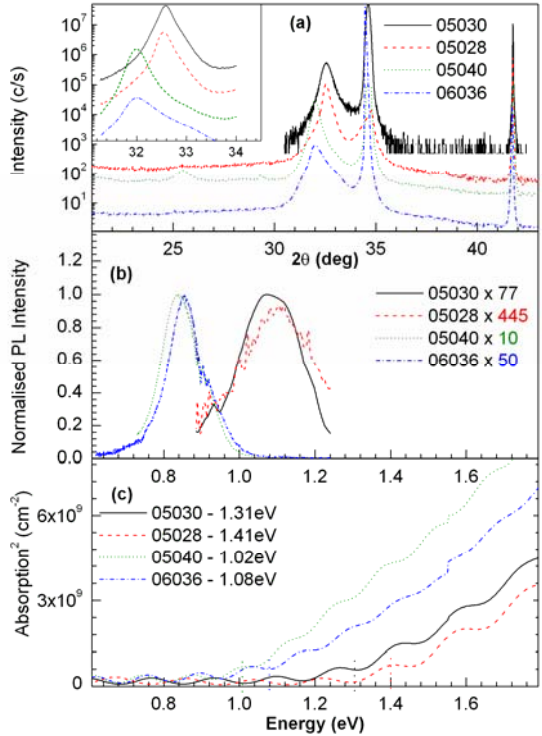

FIG. 1. (a) High resolution x-ray $\omega$-$2\theta$ scans, (b) normalized 20K photoluminescence spectra, and (c) absorption spectra of 4 selected 0.2$\mu$m-thick InN layers. The absorption edges are indicated next to the sample number. Inset in part (a) shows structure near InN (0002) peak

in two categories: (a) samples having a $2\theta$ value ~32.4° and (b) samples having a $2\theta$ value ~32°. This depends critically on the growth conditions: for example, larger $2\theta$ values (smaller lattice constant) are obtained from layers deposited at V/III ratios ~10,000 or less, while ones deposited at higher V/III ratios have smaller $2\theta$ values, corresponding to slightly larger values of lattice constants. The underlying reasons for the influence of growth parameters on the position of the InN peak are discussed separately[15]. The corresponding absorption curves near the band edge (Fig. 1(b)) and the low temperature PL spectra (Fig. 1(c)) also show that there seem to be two broad categories of samples. The absorption spectra show band edges around 1.0-1.1 eV, and around 1.3-1.4 eV for the samples having larger and smaller lattice constants respectively. This trend is also seen in the PL spectra, samples with larger lattice constants have PL peaks around 0.82 eV while samples with smaller lattice constant have weaker PL with emission centered at about 1.1eV. The latter samples are grown at a relatively low V/III ratio, not optimal for high quality InN, and the weaker luminescence possibly results from a higher concentration of defects.

The lattice parameters (*c* and *a* values) for the various layers were evaluated[12] using data from different (*hkl*) reflections. Thereafter the strain is calculated and the corresponding stress evaluated using the reported values of lattice constants of unstrained InN and the compliance tensor components[9,16]. The samples fall into two broad categories of lattice parameters and have compressive stress (a) ~ 4-5 GPa and (b) ~12-14 GPa. The presence of such a large stress in the InN layer should shift the band edge in addition to the effects due to the large carrier concentration. To further analyze the nature of stress and to eliminate other possible causes for the shift in lattice constants we have examined a range of samples grown under different conditions. This data is plotted in Fig. 2, which shows the relative shift in lattice constant from the "unstrained" literature value[13] along the *c*- and *a*-axes, for 11 samples. The dashed and solid lines show the expected changes in lattice constants for purely hydrostatic or purely biaxial strains. Points shown as open circles are InN epilayers grown directly on sapphire without a GaN buffer. For comparison, the values for MBE-grown nearly unstrained sample from two groups (data from Refs. 9,13) are also shown. From the scatter in the points it can be clearly seen that most samples lie close to the line of pure hydrostatic strain, with a relatively small biaxial component. This biaxial component increases slightly with an increase in the hydrostatic strain. While there is a small dispersion, many of the samples seem to fall into two categories, of about 1% and 3% strain respectively, consistent with the picture seen in the representative samples discussed in Fig. 1. It should be pointed out that InN layers grown directly on sapphire without a GaN buffer also fall into the two categories. This rules out InGaN alloy formation due to intermixing of gallium from the GaN buffer as the cause of any shift in InN $2\theta$ peak position. Hence, the lattice constant is determined primarily by the growth parameters for the InN layer, and not influenced by the buffer layer.

From Hall measurements, the values of the background concentration for the films whose data is plotted in Fig. 2 varies between $1.6 \times 10^{19}$cm$^{-3}$ and $4.3 \times 10^{19}$cm$^{-3}$. Assuming $E_g(0)$ of InN to be 0.7eV, and using the non-parabolic model in Ref. 4 the shift in band edge due to the Burstein-Moss shift for such a change in carrier concentration is estimated to be between 0.14eV and 0.28eV. This shift alone cannot explain the observed variation in the value of the absorption edge from ~1.0eV to 1.4eV.

The resultant stress from the presence of large (1%-3%) strains in the InN layer should shift the band edge in addition to the effects due to the large carrier concentration. To estimate the influence of stress on the electronic band structure, we have calculated the InN band edge shift using the Bir-Pikus Hamiltonian. The calculation procedure has been described in detail earlier[16], the elastic constants, deformation potentials and related parameters used are given in Ref. 17. Unstrained InN has three closely-spaced valence bands (VB) at the Brillouin zone center. The calculation accounts for strain that arises from having a mixture of hydrostatic and biaxial stress components, which in general mixes the VBs. Fig. 3 compares the values of the observed optical properties and calculated transition energy incorporating the effects of strain for a set of 7 samples of varying lattice constants. The measured absorption edge (filled circles), the values corrected for Burstein Moss shift (open circles), and PL peak position (triangles) can be compared with the data points plotted as open squares which represent the calculated transition energy involving the conduction band and the topmost strain-modified VB. While the actual value of the measured and calculated transition energies are influenced by the choice of materials parameters and do not agree exactly, the trend across the samples is quite evident. (The dashed lines are a guide to the eye). Further, the calculations do not take into account bandgap renormalization, which would slightly reduce the transition energies, and improve the agreement. From the figure, it is clear that hydrostatic stress is a key parameter that influences the shift in bandedge. This shift with stress is more prominent for the absorption edge than for the PL peak position, which is similar to that reported on studies[18] of the effect of *externally* applied hydrostatic pressure on the band edge of InN. It is worth noting that despite different growth routes, the values of the pressure coefficient of bandgap reported in Ref. 18 (3 meV/kbar (absorption)) are in the same range as values seen in our samples (~5 meV/kbar). This further strengthens our hypothesis that the internal hydrostatic stress plays an important role in determining the bandgap measured in InN epilayers.

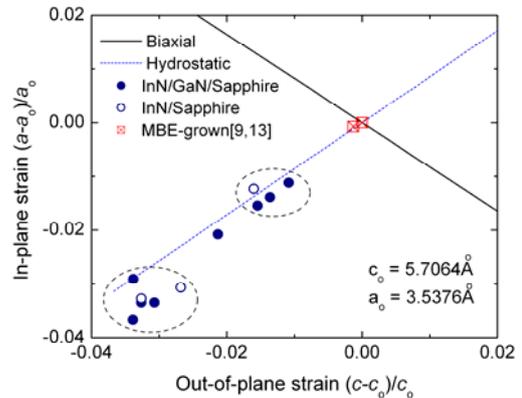

FIG. 2. Relative shift in a- and c-axis lattice parameters of InN epilayers grown under various conditions. Solid circles indicate layers grown on GaN buffers, while open circles are samples grown directly on sapphire. Lines indicate expected shifts for pure biaxial and pure hydrostatic strains.

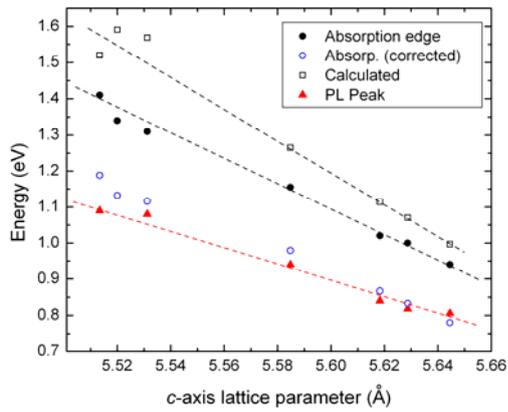

*FIG. 3. Plot of measured absorption edge (filled circles), absorption edge corrected for Burstein Moss shift (open circles), PL peak position (filled triangles), and calculated transition energy involving the conduction band and the topmost strain modified valence band (open squares) as a function of c-axis lattice constant for 7 selected InN layers. The dashed lines are a guide to the eye. See text for details.*

The knowledge of the MOVPE growth conditions that lead to relatively high- or low-stresses in the InN layer allows us to make a reasonable conjecture on the probable causes underlying this[15]. In brief, the stress is determined primarily by the nitrogen vacancies in the layer. At high V/III ratios (i.e. high ammonia flow), or at higher growth temperature (better ammonia cracking) the amount of available nitrogen species at the growth surface is higher, thus reducing the number of nitrogen vacancies, and hence the deformation of the unit cell, and consequently the stress in the layer. A similar decrease in lattice parameter, for both $c$- and $a$-axis, due to nitrogen vacancies has been reported for GaN [12,19].

In conclusion, by carefully measuring the lattice constants of MOVPE grown InN epilayers we have shown a correlation between the internal hydrostatic stress in the layer and the value of the optical absorption edge, and the PL emission wavelength. The trend observed is in reasonable agreement with calculated transition energies. We find that our MOVPE grown samples typically fall into two broad categories of stress, with resultant PL emission around 0.8eV and 1.1eV. This stress-related shift, has typically not been considered in the literature before, and may be important in determining the optical properties of InN layers.